# Lapis SOI Pixel Process

Masao Okihara[1], Hiroki Kasai[2], Noriyuki Miura[2],
Naoya Kuriyama[2], Yoshiki Nagatomo[1]

[1]LAPIS Semiconductor Co., Ltd., [2]LAPIS Semiconductor Miyagi Co., Ltd.,

0.2 um fully-depleted SOI technology has been developed a for X-ray pixel detectors. To improve the detector performance, some advanced process technologies are developing continuously.

To utilize the high resistivity FZ-SOI, slow ramp up and ramp down recipes are applied for the thermal processes in both of SOI wafer fabrication and pixel detector process. The suitable backside treatment is also applied to prevent increase of leakage current at backside damaged layer in the case of full depletion of substrate. Large detector chip about 66mm width and 30mm height can be obtained by stitching exposure technique for large detector chip. To improve cross-talk and radiation tolerance, the nested well structure and double- SOI wafer are now under investigation for advanced pixel structure.

# 1 Introduction

0.2 um fully-depleted (FD) SOI technology has been developed a for X-ray pixel detectors [1]. By using FD-SOI technology, electrical circuitry in the top silicon layer and X-ray sensor p-n junction diode in high resistivity handle wafer can be made simultaneously. It means monolithic X-ray pixel detector device can be obtained easily. FD-SOI has a promising structure to realize high performance and reliable monolithic X-ray pixel detectors.

# 2 Observations

In order to further improve the SOI pixel process, some advanced or modified process have been developing such as

(1) Suppression of back-gate effect by using buried P-well,
(2) Improvement of radiation hardness by using double-SOI wafer,
(3) High resistivity FZ-SOI process improvement,
(4) Backside treatment for fully depleted substrate,
(5) Stitching exposure for large sensor chip.

(1) Suppression of Back Gate Effect

In X-ray sensor, "large" back gate voltage is necessary to operate wider energy range of X-ray detection. Threshold voltage drop is observed with large back gate bias because of back channel turn-on. Suppression of back gate effect is indispensable to achieve monolithic FD-SOI X-ray Sensor. Back gate effect has been drastically suppressed by introducing BPW (Buried P-Well) biased with GND under the Transistors.

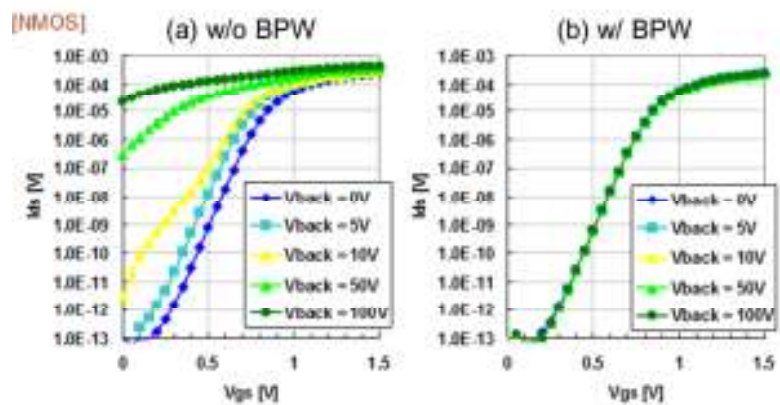

Fig.1  Back-bias dependence of IV Characteristics of NMOS w/ and w/o BPW.

Figure 1 show the IV characteristics with and without BPW. There is no Id-Vg change even with Vback=100V is observed for with BPW structure.

(2)Double SOI

Further issue is radiation effect from TID. If the BOX is charged by X-ray radiation, backside channel will be turned on like back gate effect. In order to improve radiation hardness, double-SOI wafer with middle-SOI structure has been evaluated with KEK and other collaborator. Figure 2 shows the negative view image of STEM cross-section. This middle-SOI layer, which is surrounded by red dash line, exists under the transistors. The middle-SOI can be used for compensation of BOX charging effect and also reduce the crosstalk between sensors and circuits.

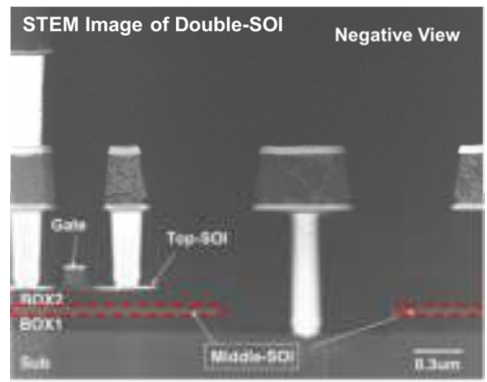

Fig. 2  STEM image of Double-SOI Structure (Negative view)

Figure 3 show IV curves as the function of radiation dose. The left one is 0V applied to middle SOI, center one is applied -2V and right one is -5V. In the case of 0V applied, drastic sub-threshold shift is observed. However, you can see the shift can be compensated with applying negative voltage to middle SOI, and IV characteristics become nearly pre-irradiation level.

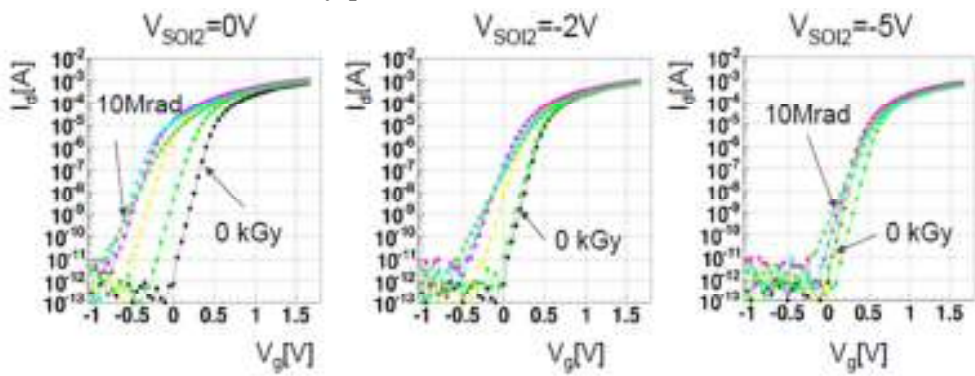

Fig.3  IV curves from Double-SOI Structure (Obtained by Univ. of Tsukuba).

(3) High Resistivity FZ-SOI Process Improvement

In order to improve the detection efficiency, high resistivity wafer is required for high X-ray energy range. CZ wafer is difficult to obtain high resistance because dissolved Oxygen turn to thermal donor in the low temperature annealing. On the other hand, FZ wafer is ideal for high-R application for its low Oxygen density since FZ method is using highly controlled atmosphere. However, one of the problems to use FZ wafers is weak mechanical strength at high temperature processes. Conventional bonded SOI process shows many slip lines were generated in the handle wafer. By optimizing process parameters, mainly decrease ramp up and ramp down rate, slip generation can be reduced to the acceptance level for pixel sensor fabrication in Fig. 4.

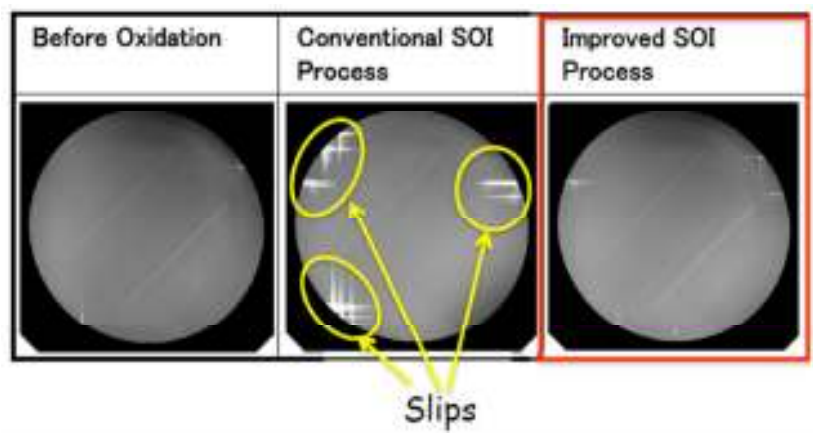

Fig. 4　X-ray Topography of FZ-SOI wafers

(4) Backside Treatment

Next topic is backside treatment to make Low Resistivity Contact and prevent Abnormal Leakage Current. In order to obtain such contact, the wet etching after back grinding and laser anneal are introduced. Figure 5 show cross-sectional backside surface with and without wet etching. The 0.3um-thick damaged layer can be removed by wet etching. In this case, it is confirmed the abnormal leakage current can be suppressed.

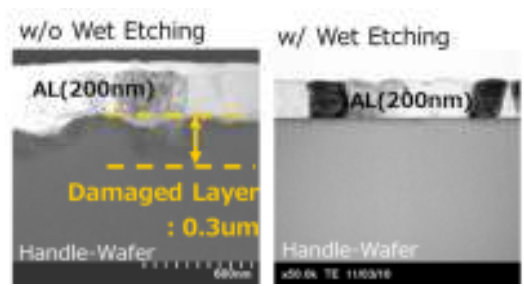

Fig. 5　TEM image of backside surface w/ and w/o wet etching

(5)Stitching Exposure technique

In order to create large sensor chip, it needs stitching some shots to make one large chip. This figure shows the shot image for normal SOI Pixel MPW run. Usually, there are lots of chips in the one shot. Fig. 6 is the stitching image for Riken's SOPHIAS chip [2]. In the SOPHIAS case, 3 sensor shots and 2 guard ring are stitched together.

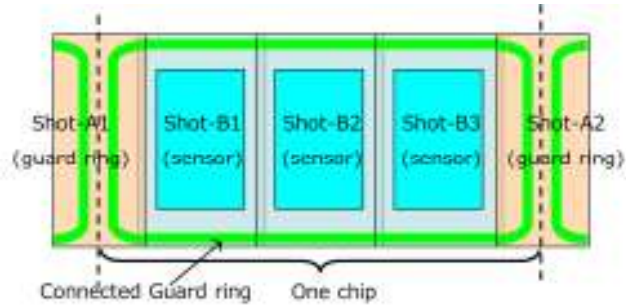

Fig. 6  Stitching image of SOPHIAS chip

There is 14um-thick boundary area between stitched shots. It was confirmed that pattern thinning in double exposure area are is less than 0.1um and Y-direction alignment between shots is about 0.015um. Also, there is no pattern loss or deformation at the stitching area.

# 3 Conclusions

SOI Pixel technology for X-ray sensor has been developed with KEK, Riken and other collaborators.

In order to improve Pixel Sensor, some process technologies have been developing or modifying so far. Such as BPW, Double-SOI, FZ-SOI Process, Backside Treatment and Stitching Exposure. Further improvements of process technology for SOI Pixel sensor are under consideration.

We have been developing several kinds of detectors with KEK, such as integration-type pixel (INTPIX), counting-type pixel (CNTPIX) and dual-mode integration type pixel (DPIX) [3-6]. Furthermore, SOI photon imaging array sensor (SOPHIAS) which has large chip aria and wide dynamic range pixel detector is developing with RIKEN team.